\documentclass[a4paper,twocolumn,showpacs,pre]{revtex4-1}
\usepackage{amsmath,amssymb}
\usepackage{graphicx,color}
\begin{document}
\title{Onset of Localization in Heterogeneous Interfacial Failure}
\author{Arne Stormo}
\email{arne.stormo@gmail.com}
\affiliation{Department of Physics, Norwegian University of Science and 
Technology, N--7491 Trondheim, Norway}

\author{Knut Skogstrand Gjerden}
\email{knut.skogstrand.gjerden@gmail.com}
\affiliation{Department of Physics, Norwegian University of Science and 
Technology, N--7491 Trondheim, Norway}

\author{Alex Hansen}
\email{Alex.Hansen@ntnu.no}
\affiliation{Department of Physics, Norwegian University of Science and 
Technology, N--7491 Trondheim, Norway}
\date{\today}
\begin{abstract}
We study numerically the failure of an interface joining two elastic materials 
under load using a fiber bundle model connected to an elastic half space. We
find that the breakdown process follows the equal load sharing fiber bundle
model without any detectable spatial correlations between the positions of
the failing fibers until localization sets in.  The onset of localization is
an instability, not a phase transition.  Depending on the elastic constant
describing the elastic half space, localization sets in before or after
the critical load causing the interface to fail completely, is reached.  There
is a crossover between failure due to localization or failure without spatial
correlations when tuning the elastic constant, not a phase transition.  
Contrary to earlier claims based on models different from ours, we find that 
a finite fraction of fibers must fail before the critical load is attained, 
even in the extreme localization regime, i.e.\ for very small elastic 
constant. We furthermore find that the critical load remains finite for all
values of the elastic constant in the limit of an infinitely large system.   
\end{abstract}
\maketitle

The joining of interfaces, e.g.\ by welding or gluing, is an important
part of everyday technology, a technology that has been refined through
the centuries.  When joined interfaces are subject to excessive loads, 
failure occurs. Often it is not the joints themselves that fail, but
the material that surrounds them, as the joints themselves are the 
stronger.

The aim of this work is, however, not to study failure of such joints
with improvement of technology in mind.  Rather, we take the point of
view that failure of a heterogeneous joined interface provides a 
simplified model for fracture in bulk materials.  Such an idea is not
new.  Schmittbuhl et al.\ \cite{srvm95} and Schmittbuhl and M{\aa}l{\o}y 
\cite{sm97} studied, first computationally, then experimentally,
the roughness of the fracture front moving through a sintered interface
between two Plexiglas plates that are being plied apart in a mode-I 
fashion.  The study of the fluctuations of this fracture front provides 
much insight into the much more complex morphology of three-dimensional
fracture surfaces \cite{bb11}.  

We focus on the phenomenon of {\it localization\/} in this work.  Local 
failure occurs either because the material is weaker at that spot or because
it is more loaded there than elsewhere.  Differences in local strength
is due to heterogeneities.  Differences in loading is due to structure in
the stress field.  If we assume that the interface is loaded uniformly ---
the local stress field will be quite uniform.  Local failure will occur
because of material weakness.  For the simple reason that the further
away we search from a point where a failure has occurred, the weaker the
weakest spot we have found so far will be, localization is disfavored.      
Heterogeneity in strength induces a ``repulsion" between the local failures.  
However, when failed areas build up, local stress is concentrated at the rim 
of the failed areas making these regions liable to fail.  Heterogeneity in
the stress field induces an ``attraction" between local failures.  

Localization occurs when attraction wins over repulsion.  A transition
in the failure process occurs at this point.  What is the nature of this
transition?  As we shall demonstrate, it is not a phase transition, but
a crossover phenomenon.  

When localization sets in immediately in the breakdown process, it is
normally expected that the system is infinitely fragile in the limit of
infinitely large system: As soon as a single fiber breaks at a given load, 
the entire system breaks down at that load \cite{phc10}. As we will 
demonstrate, this is {\it not\/} the case here.

We base our work on the discretized model for interfacial failure proposed by 
Batrouni et al.\ \cite{bhs02}. A square array of $L\times L=N$ 
linearly elastic fibers placed a distance $a$ apart connects a stiff 
half-space with a linearly elastic half space characterized by a Young 
modulus $E$ and a Poisson ratio $\nu$ (which we assume a typical value of $0.25$ in the following). 
Each fiber, indexed by $i$, 
has an elastic constant $k$ and fails irreversibly if it is elongated beyond 
an individual threshold value $t_i$.  The threshold values are drawn from a
uniform distribution on the unit interval.      

The separation of the two half spaces are controlled by displacing the hard
medium by a distance $D$ orthogonal to the interface where the fibers sit.
Fiber $i$ then experiences a force
\begin{equation}
\label{eq1}
f_i=-k(u_i-D)\;,
\end{equation}
where $u_i$ is the local displacement of the softer half space at the 
position of fiber $i$.  The forces 
from the fibers are transmitted through the softer elastic 
medium via the Green function \cite{l29,j85,ll86}
\begin{equation}
\label{greensum}
u_i = \sum_jG_{ij}f_i\;,
\end{equation}
where
\begin{equation}
\label{green}
G^{}_{ij}= \frac{1-\nu^2}{\pi E a^2} \iint^{a/2}_{-a/2}
\frac{dx'dy'}{|\vec{r}_i(x,y)-\vec{r}_j(x',y')|}\;. 
\end{equation}
$\vec{r}_i-\vec{r}_j$ denotes the distance between fibers 
$i$ and $j$ at positions $\vec{r}_i$ and $\vec{r}_j$ respectively. 

The Green function (\ref{green}) is modified by the presence of boundaries
due to the finite size  $(La)\times (La) = N a^2$ of the interface. We
assume periodic boundary conditions and take into account the first 
reflected images.

We note that if distances are measured in units of $a$, the 
Green function (\ref{green}) is proportional to $(Ea)^{-1}$. Likewise, from
Eq.\ (\ref{eq1}), we see that the elastic constant of the fibers, $k$ must be
proportional to $a^2$.  Hence, if we change the linear size of the system, 
$(La)\to \lambda (La)$, while keeping the discretization $a$ fixed, we change 
only $L\to \lambda L$ in the model whereas we keep the parameters $Ea$ and
$k$ fixed.  On the other hand, if we change the 
discretization $a\to a/\lambda$ while leaving the size of the system fixed, 
we change $L\to \lambda L$ and the parameters $(Ea)\to (Ea)\lambda$ and
$k\to k/\lambda^2$.    

A given fiber breaks irreversibly (its elastic constant is
set to zero) if stretched beyond a threshold value assigned from a 
spatially uncorrelated probability distribution.  We choose the simplest:
a uniform distribution.  The model is quasi-static, 
and in lieu of time, we measure the fraction of fibers that have broken,
denoted by $p$. The load carried by the system 
is $\sigma(p)=\sum_i f_i/N$, and when $\sigma$ reaches its 
maximum, any extra load will result in a complete catastrophic failure.
We denote this the \textit{critical load}, $\sigma_c$, 
and the corresponding $p_c$, the \textit{failure point}.

In the limit of $(Ea/L)\to\infty$, the model becomes identical to the 
equal load sharing fiber bundle model (ELS) \cite{phc10,p26,d45}.  
On the other hand, for
small values of $(Ea/L)$, it does not approach any existing models.  Models
do exist, e.g.\ the local load sharing fiber bundle model (LLS) \cite{hp81}, 
where the nearest surviving fibers absorbs the entire load that a fiber 
was carrying when failing. Another model, introduced by Hidalgo et 
al.\ \cite{hmkh02}, distributes the added load around a failed fiber as a 
power law in the distance from the failed fiber.  In both models, there is no 
elastic response by the planes defining the interface.        

We have studied systems of size $L=256$,  $L=128$,  $L=64$, $L=32$ $L=16$ and $L=8$ with 
$10$, $100$, $1000$, $10000$, $10000$ and $10000$ samples respectively.  We explore a range
of elastic constants $e \equiv (Ea/L)$ in the range $e_{\text{soft}} \le
e \le e_{\text{stiff}}$ where $e_{\text{soft}}=2^{-17}=7.63\times 10^{-6}$ and 
$e_{\text{stiff}} =2^6=32$. 

\begin{figure}
\includegraphics[scale=0.79]{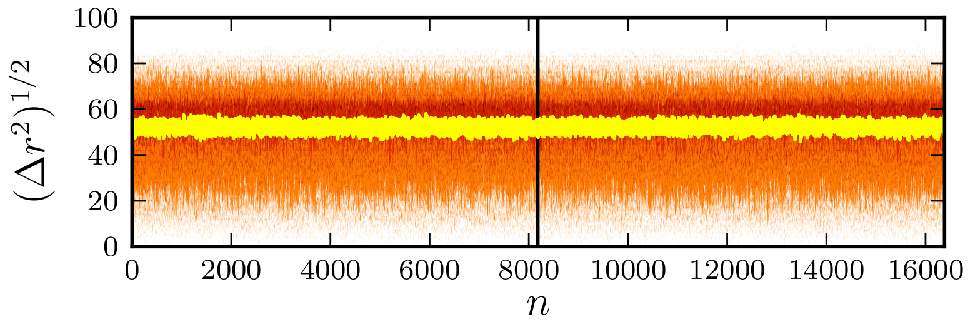}
\includegraphics[scale=0.79]{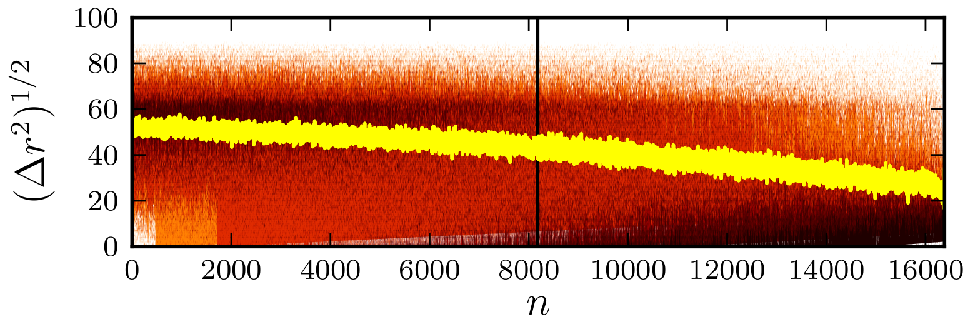}
\includegraphics[scale=0.79]{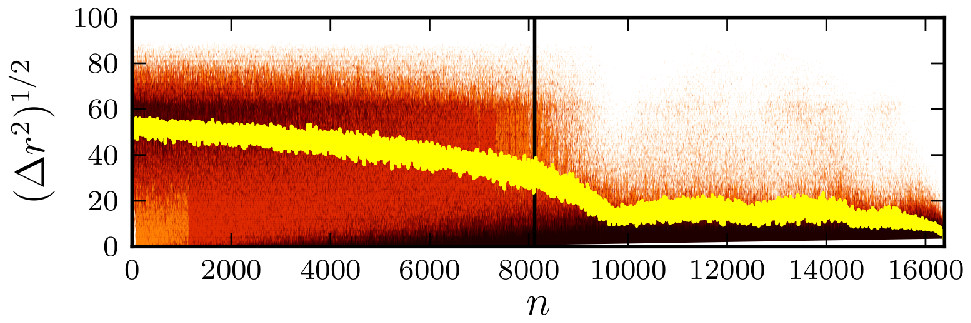}
\includegraphics[scale=0.79]{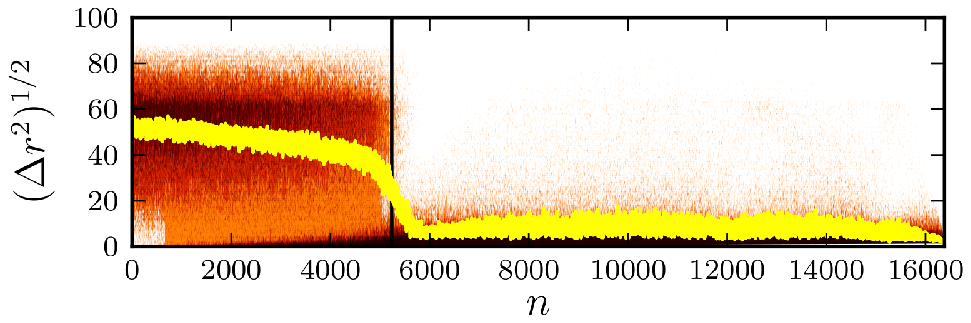}

\caption{(Color online.) Histogram over the distance between consecutively failing fibers, 
$(\Delta r^2(n))^{1/2}$, as function of the number of failed fibers $n=pN$. 
Darker colors correspond to higher density. The vertical bar indicates the 
failure point, $p_cN$. The curve shows the running average 
$\langle\Delta r^2(n)\rangle^{1/2}$. In all figures, $L=128$. 100 simulations  
was used to construct the histogram.
From top to bottom, $e=32$, $e=2^{-3.678}=0.0781$, $e=2^{-6}=0.0156$ and 
$e=2^{-17}=7.63\cdot 10^{-6}$.} 
\label{fig1}
\end{figure}

In order to visualize localization, we record the square distance between 
consecutively failing fibers, $\Delta r^2$.  If the positions of the failing are
completely random, as is the case in the ELS fiber bundle model, the average
distance is $\langle \Delta r^2\rangle^{1/2}=L/\sqrt{6} \approx 0.408 L$.  We show in 
Fig.\ \ref{fig1} a succession of histograms of $(\Delta r^2)^{1/2}$.  That is, we

record $\Delta r^2(n)$ as a function of the number of failed fibers, $n=pN$ ---
our ``time" parameter. We then sum the number of times $\Delta r(n)$ has
had a particular value $(\Delta r^2)^{1/2}$ at $n$ for several independent simulations,
hence creating a histogram for each $n$. Darker colors signifies more hits at 
that value of $(\Delta r)^{1/2}$.  The curve shows
the average value $\langle \Delta r^2(n)\rangle^{1/2}$.  With $L=128$, we see that
for the four different elastic constants $e$ that we show, $e=32$, $e=2^{-3.678}$ 
(this value is chosen to make the figure comparable to the results in Batrouni et al.\ \cite{bhs02}, 
where $L=128$, $E=10$, and $a=1$, which gives $e=0.0781=2^{-3.678}$.)
$e=2^{-6}$ and $e=2^{-17}$, $\langle\Delta r^2(n)\rangle^{1/2}$ starts out being
close to the ELF fiber bundle model value 52.26.  
The vertical line in each figure shows the failure point $n_c = p_c N$.  

Rising the value if the elastic constant $e$ above $e_{\text{stiff}}$ 
or below $e_{\text{soft}}$, will not lead to changes from the uppermost and 
lowermost panels in Fig.\ \ref{fig1}.  At the highest value of $e$, the 
system behaves as the ELS fiber bundle model throughout the entire breakdown
process: The average distance between consecutively failing fibers, 
$\langle\Delta r^2(n)\rangle^{1/2}$ remains constant throughout the process.  The ELS fiber
bundle model predicts $p_c=1/2$ so that $n_c=8192$.       

As the system gets softer, both $\langle\Delta r^2(n)\rangle^{1/2}$ and $n_c$ decrease.
We see in three lower panels in Fig.\ \ref{fig1} that there is an abrupt
change in $\langle\Delta r^2(n)\rangle^{1/2}$ for some range of $n$ values.  This
is localization.  We also see that the failure point does {\it not\/} fall
to zero as $e$ is lowered.  Even for the smallest value in Fig.\ \ref{fig1},
$e=e_{\text{soft}}=2^{-17}$, $n_c$ is significantly different from zero. 

\begin{figure}
\includegraphics[scale=0.7]{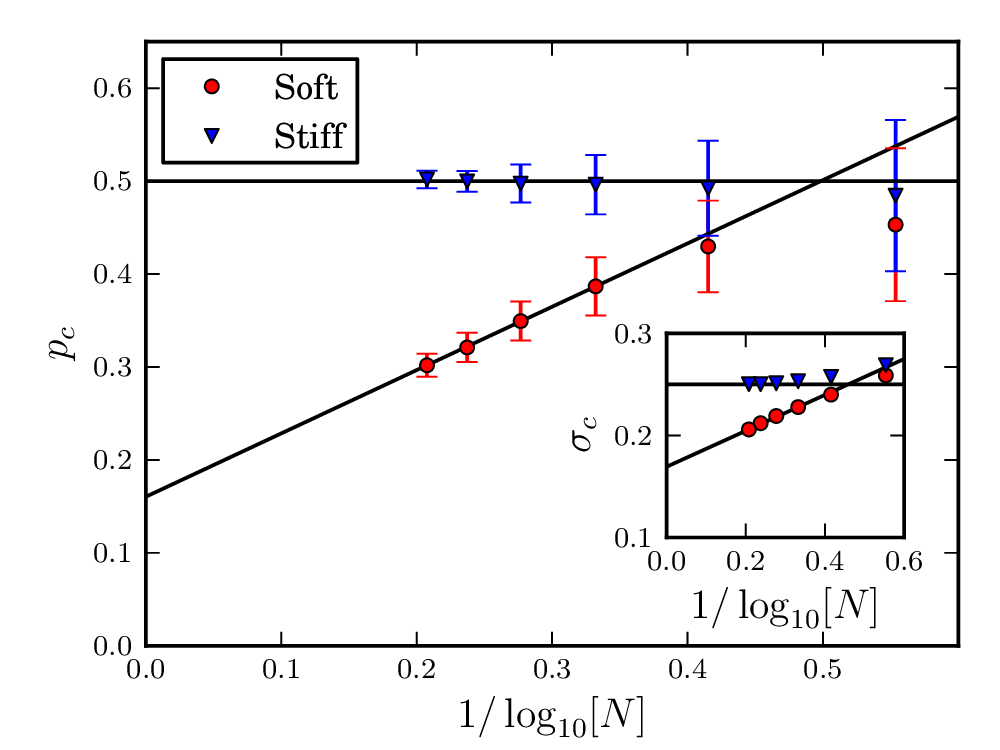}
\caption{(Color online.) Finite size analysis of failure point and 
critical loading. Both $p_c$ (main plot) and 
$\sigma_c$ (encapsulated) is plotted against 
$1/\log_{10}[N]$, for $N=8^2=256$ to $N=256^2=65536$ and both have finite values for any $N$. 
The slope of the soft systems are $\alpha_{p_c}=0.68$ and 
$\alpha_{\sigma_c}=0.16$.}
\label{fig2}
\end{figure}

Fig.\ \ref{fig2} shows the failure point $p_c$ as a function of the
inverse of the logarithm of total number of fibers, $1/\log_{10}(N)$. 
From this figure, we may
extrapolate the value of $p_c$ in the limit of infinitely large system. 
We find $p_c(N\to\infty)=p_c^\infty=0.16$ when $e=e_{\text{soft}}$ and
$p_c^\infty=0.5$ when $e=e_{\text{stiff}}$.  Likewise, we may extrapolate
the critical load $\sigma_c$ --- see the insert in the figure.  We find
through extrapolation that $\sigma_c^\infty=0.17$ for $e=e_{\text{soft}}$ 
and $\sigma_c^\infty=0.25$ for $e=e_{\text{stiff}}$, the value expected for 
the ELS fiber bundle model.    

It is a surprising result that neither $p_c^\infty$ nor $\sigma_c^\infty$ are
zero for small value of $e$.  The LLS fiber bundle model predicts that 
$\sigma_c \sim 1/\log_{10}(N)$ with $\sigma_c^\infty=0$ \cite{zd95,khh97}.  Hidalgo 
et al.\ \cite{hmkh02} present numerical evidence that their model also has
$\sigma_c^\infty=0$.  Hence, in both of these models, $p_c^\infty=0$. 

Batrouni et al.\ \cite{bhs02} studied the structure of the clusters 
of failed fibers at $p_c$, claiming that at the failure point, they are 
distributed according to a power law with exponent -1.6.  This would
indicate a critical point at $p=p_c$.  

\begin{figure}
\includegraphics[scale=0.7]{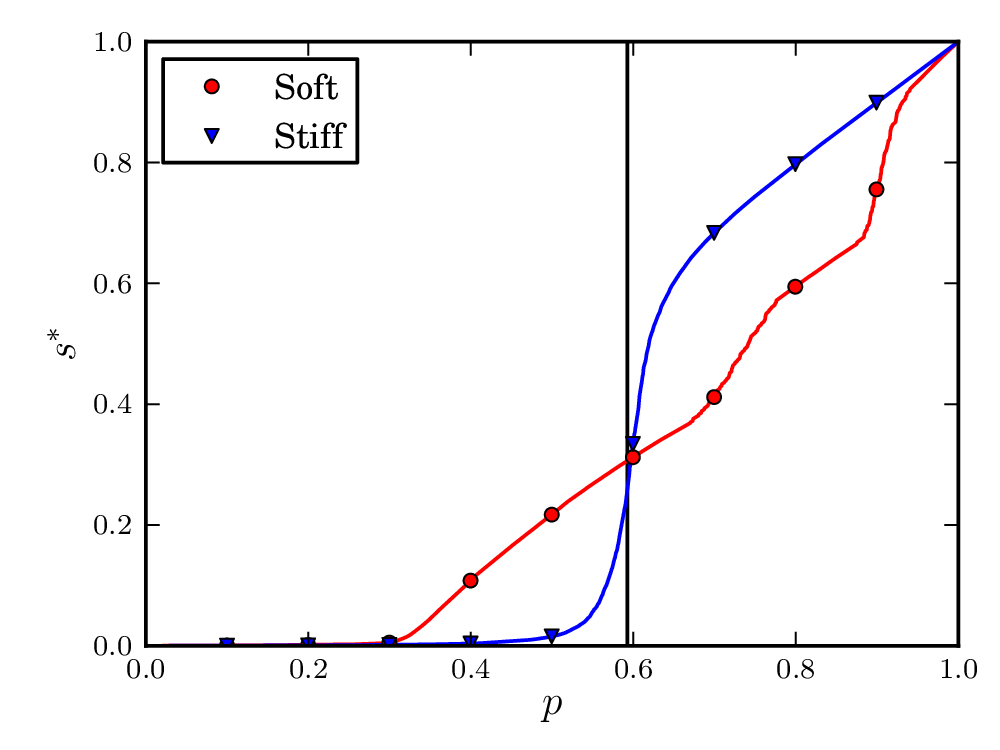}
\caption{(Color online.) Density of the largest cluster of failed fibers, $s^*$, as
a function of the damage $p$ for $e=e_{\text{stiff}}$ and $e=e_{\text{soft}}$.
Here $L=128$. The vertical bar indicates the percolation point $p=0.59274$.}
\label{fig3}
\end{figure}

Referring to Fig.\ \ref{fig1}, we see 
that for $e=e_{\text{stiff}}=32$, the system behaves as the ELS fiber bundle 
model where the position of the fibers that fail bear no correlations 
among themselves. It is clear when observing the average distance between 
consecutively failing fibers, $\langle\Delta r(n)^2\rangle^{1/2}$ which 
essentially remains close to the ELS fiber bundle, where 
$\langle\Delta r^2(n)\rangle^{1/2} = L/\sqrt{6}$.  Hence, we
expect that the clusters follow {\it percolation\/} theory \cite{sa94}.  In
Fig.\ \ref{fig3} we show the density of of the largest cluster of failed fibers,
$s^*$, as a function of $p$ for $e=e_{\text{stiff}}$ and $e=e_{\text{soft}}$.  
In the case of the soft system, we see that at $p\approx 0.25$, the largest
cluster becomes visible and grows essentially linearly with $p$.  This behavior is due to localization. 
When $p$ approaches 1, there are jumps in $s^*$, because of coalescence of clusters.
On the other hand, when $e=e_{\text{stiff}}$, we see behavior consistent with 
percolation theory.  When $p$ is in the vicinity of $p=0.59274$, the site
percolation threshold on the square lattice \cite{fdb08}, $s^*$ shoots up
and thereafter evolve linearly in $p$.  

\begin{figure}
\includegraphics[scale=0.55]{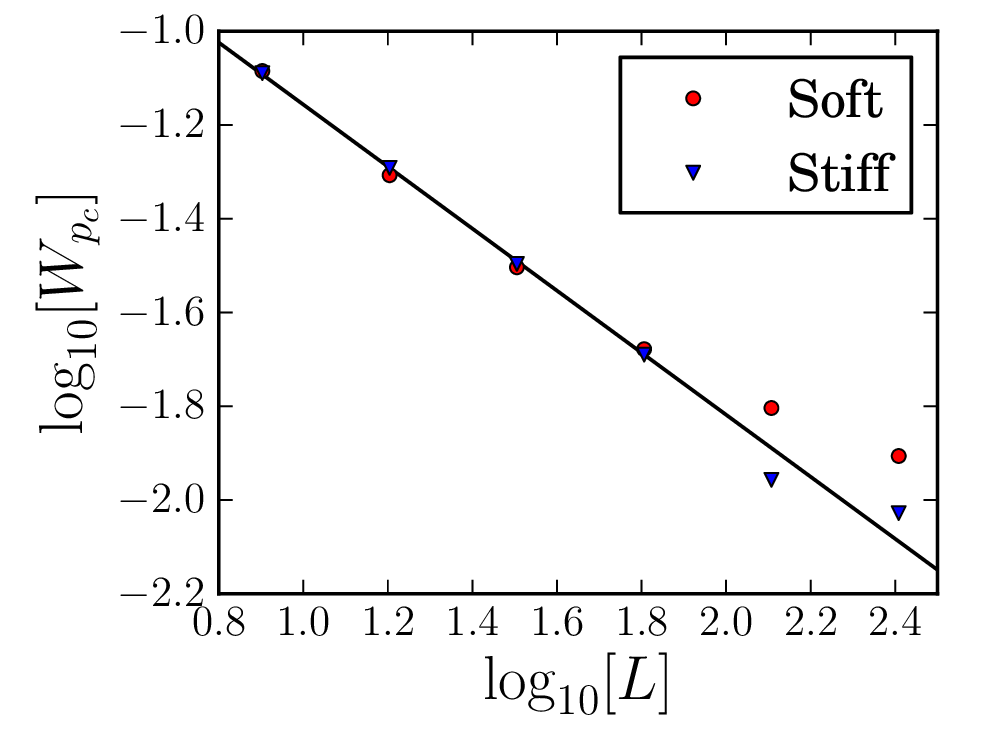}
\caption{(Color online.) Fluctuations of $p_c$, $W_c=\sqrt{\langle p_c^2\rangle-\langle p_c\rangle^2}$ plotted against $L$ for $L=8$ to $L=256$. 
The slope of the black line is $-2/3$. The stiff system follows the slope, while the soft systems deviates.}
\label{fig4}
\end{figure}

The failure point at which the system fails catastrophically, $p_c$,
occurs long before the jump in $s^*$ for $e=e_{\text{stiff}}$.  This
is an indication that the system is {\it not\/} critical at the failure
point.  Schmittbuhl et al.\ \cite{shb03} measured the fluctuations
of the failure point as a function of the system size, finding 
$\Delta p_c \sim 1/L^{0.65}$. This is consistent with the GLS fiber
bundle model.  Daniels and Skyrme \cite{ds89} showed that the statistical
distribution of the critical elongation in the GLS fiber bundle model
has has the form 
\begin{equation}
\label{fu}
\rho(u_c)du_c=N^{1/3}f[CN^{1/3}(u_c-\langle u_c\rangle)]du_c\;,
\end{equation} 

$C$ is a constant only dependent on the threshold distribution and $u_c$ is
the critical elongation.  This leads immediately to 
\begin{equation}
\label{deltau}
\Delta p_c \sim \langle(u_c-\langle u_c\rangle)^2\rangle^{1/2} 
\sim N^{-1/3} = L^{-2/3}\;,
\end{equation}
where we have used the assumption that the threshold distribution is 
uniform in the vicinity of $\langle u_c\rangle$ in relating $p_c$ to $u_c$.
From Fig.\ \ref{fig4} we can see that the stiff system scales as $L^{-2/3}$, while the soft systems deviates. 
We conclude that we cannot detect spatial correlation in the failure
process beyond an uncorrelated percolation process for $e=e_{\text{stiff}}$.

\begin{figure}
\includegraphics[scale=0.7]{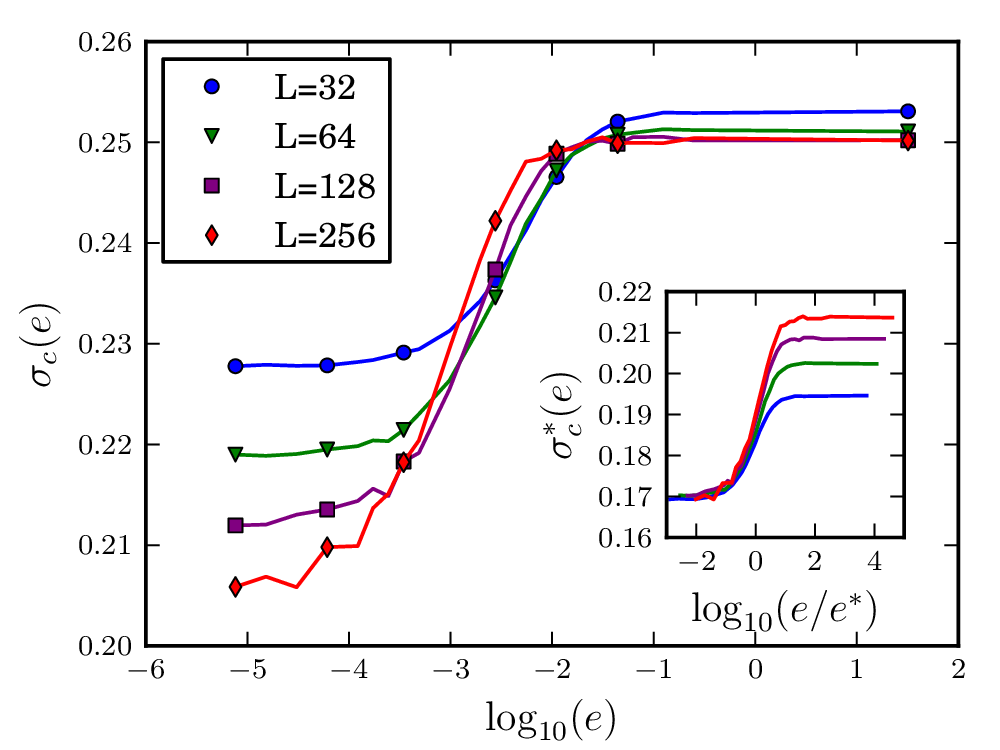}
\caption{(Color online.) The critical load, $\sigma_c$, plotted 
against $\log_{10}(e)$ for several system sizes. 
The insert shows the rescaled $\sigma_c^*$ is plotted against 
$ \log_{10}(e/e^*)$, demonstrating data collapse.}
\label{fig5}
\end{figure}

We now proceed to study the onset of localization. Fig.\ \ref{fig5} shows 
the critical load $\sigma_c$ as a function of $e$ 
for systems of size $L=32$ to $L=256$. As $e$ decreases, we observe that 
$\sigma_c$ drops from the ELS fiber bundle value, goes through a crossover 
and ends in a stable $\sigma_c$ for each $L$ in the soft regime.  We define 
$e^*$ by setting
\begin{equation}
\label{estar}
\sigma_c(e^*)=\frac{1}{2}
\left(\sigma_c(e_{\text{stiff}})+\sigma_c(e_{\text{soft}})\right)\;.
\end{equation} 
We then define
\begin{equation}
\label{sigmastar}
\sigma_c^*(e)=
\sigma_c(e)-\frac{\alpha_{\sigma_c}}{\log_{10}[N]}\;.
\end{equation}

We show $\sigma_c^*(e)$ vs.\ $\log_{10}(e/e^*)$ in the insert in Fig.\ \ref{fig5}.
As the correction term $-\alpha_{\sigma_c}/\log_{10}[N]\rightarrow 0$ in the macroscopic limit, 
we know that the curve never grow past $\sigma_c(e^*)=0.25$. The largest gradient in 
$\sigma_c{e}$ seem to converge around 
\begin{equation}
\label{sigmagrad}
 \frac{\Delta \sigma_c(e)}{\Delta\log_{10}[e/e^*]}=0.03,
\end{equation} 
and the shape of the curve is kept. 
The onset of localization is 
{\it not\/} a phase transition, but a {\it crossover:\/} there are no 
divergences anywhere in the derivative of this curve.  

The following picture then emerges: For a given elastic constant, $e$, the 
breakdown process starts out as described by the ELS fiber bundle model.  The 
spatial correlations between the failing fibers seems to be so weak that 
it can be described as an uncorrelated percolation process.  If the elastic
constant $e$ is large enough, the system will undergo both the ELS fiber 
bundle model failure point and the percolation transition.  Depending on the 
threshold distribution, the ordering of the two events, the ELS failure point
and the percolation transition, may be reversed. With lower 
elastic constant, localization sets in and breaks off the ELS fiber bundle 
breakdown process.  When localization sets in, all failure activity is then
essentially limited to the rim of a growing cluster of failed fibers.  The
onset of localization is an instability, not a phase transition.

We thank M.\ Gr{\o}va, S.\ Pradhan, S.\ Sinha and B.\ Skjetne for useful discussions.  This work
was partially supported by the Norwegian Research Council through 
grant no.\ 177591.  We thank NOTUR for allocation of computer time.


\end{document}